\documentclass[preprint]{aastex}
\usepackage{mkfig}

\begin{document}

\title{Sequential Small Coronal Mass Ejections Observed In~situ and in
  White-Light Images by Parker Solar Probe}

\author{Brian E. Wood\altaffilmark{1}, Phillip Hess\altaffilmark{1},
  Yu Chen\altaffilmark{2}, Qiang Hu\altaffilmark{2,3}}
\altaffiltext{1}{Naval Research Laboratory, Space Science Division,
  Washington, DC 20375, USA; brian.wood@nrl.navy.mil}
\altaffiltext{2}{Center for Space Plasma and Aeronomic Research (CSPAR),
  The University of Alabama in Huntsville,  Huntsville, AL 35805, USA}
\altaffiltext{3}{Department of Space Science,
  The University of Alabama in Huntsville,  Huntsville, AL 35805, USA}

\begin{abstract}

     We reconstruct the morphology and kinematics of a series of small
transients that erupt from the Sun on 2021~April~24 using observations
primarily from Parker Solar Probe (PSP).  These sequential small coronal
mass ejections (CMEs) may be the product of continuous reconnection
at a current sheet, a macroscopic example of the more microscopic
reconnection activity that has been proposed to accelerate the solar
wind more generally.  These particular CMEs are of interest because
they are the first CMEs to hit PSP and be simultaneously imaged by it,
using the Wide-field Imager for Solar Probe (WISPR) instrument.
Based on imaging from WISPR and STEREO-A, we identify and model
six discrete transients, and determine that it is the second of
them (CME2) that first hits PSP, although PSP later more obliquely
encounters the third transient as well.  Signatures of these
encounters are seen in the PSP in~situ data.  Within these data,
we identify six candidate magnetic flux ropes (MFRs),
all but one of which are associated with the second transient.
The five CME2 MFRs have orientations roughly consistent
with PSP encountering the right sides of roughly E-W oriented MFRs,
which are sloping back towards the Sun.

\end{abstract}


\section{Introduction}

     Parker Solar Probe (PSP) provides an opportunity to study
coronal mass ejections (CMEs) closer to the Sun than ever before.
However, the fraction of time PSP spends close to the Sun in its
highly elliptical orbit is relatively small, and the odds of
being hit by a CME during a particular perihelion
passage are therefore relatively low.
Thus, there are so far a limited number of events available for study
during the 15 orbits that PSP has made around the Sun as of 2023 April.
The most well-studied event dates back to the very first perihelion
passage, where on 2018 November 11 shortly after perihelion (2018 November 6)
PSP encountered a CME while at a distance of 53~R$_{\odot}$ from the Sun
\citep{kek20,tnc20,swg20}.
Other events observed in situ by PSP farther from the Sun have been
analyzed by \citet{rmw21}, \citet{qh22}, and \citet{cm22}.

     The PSP mission is primarily designed to study the solar wind and
CMEs using a suite of on board particle and field detectors.
However, PSP also possesses a white light imaging instrument,
the Wide-field Imager
for Solar Probe (WISPR) \citep{av16}.  The two heliospheric
imagers that consistitute PSP/WISPR observe the solar wind in the
ram direction of PSP's orbit, and are typically taking data only for
about a month surrounding each perihelion.  A number of CMEs have
been observed by WISPR during these periods, dating back to the first orbit,
with a small circular CME observed on 2018 November 1 \citep{ph20,apr20},
followed by a ``streamer blob'' event seen
very near perihelion on 2018 November 5 \citep{bew20}.  The
transients studied so far have tended to be rather small and slow, but these
images have still allowed CME structures to be seen in much greater
detail than ever before \citep{pcl20,pcl21,bew21,crb22,rah22}.

     Given that WISPR observes in the direction of the spacecraft
motion, WISPR
sees transients passing in front of the spacecraft in its orbit, which
are typically not hitting PSP.  We report here on the first CME
that both directly hits PSP and is imaged by WISPR while doing so, offering
the first opportunity for WISPR and PSP's in situ instruments to be used to
study the same CME.  Unfortunately, this CME encounter is a complicated
one, with WISPR observing many fronts passing through the field of view on
2021 April 24-25, just before PSP's eighth perihelion on April 29, with
the spacecraft 46 R$_{\odot}$ from the Sun.  Necessary clarity is provided by
coronagraphic observations from STEREO-A, showing that the eruptive activity
is more accurately described as a series of small CMEs erupting off the east
limb, as viewed from STEREO-A's perspective.  A central task of our analysis
will be to utilize the WISPR and STEREO-A images to determine which of
the small transients are actually hitting PSP, a necessary step toward
properly interpreting the in~situ signatures of the activity observed
by PSP.  There is also a prominence eruption associated with this
activity, observed in EUV images from both STEREO-A and Solar Orbiter, but
we refer the reader to \citet{tn23} for a detailed discussion of the
EUV data.

\section{Observations}

\begin{figure}[t]
\plotfiddle{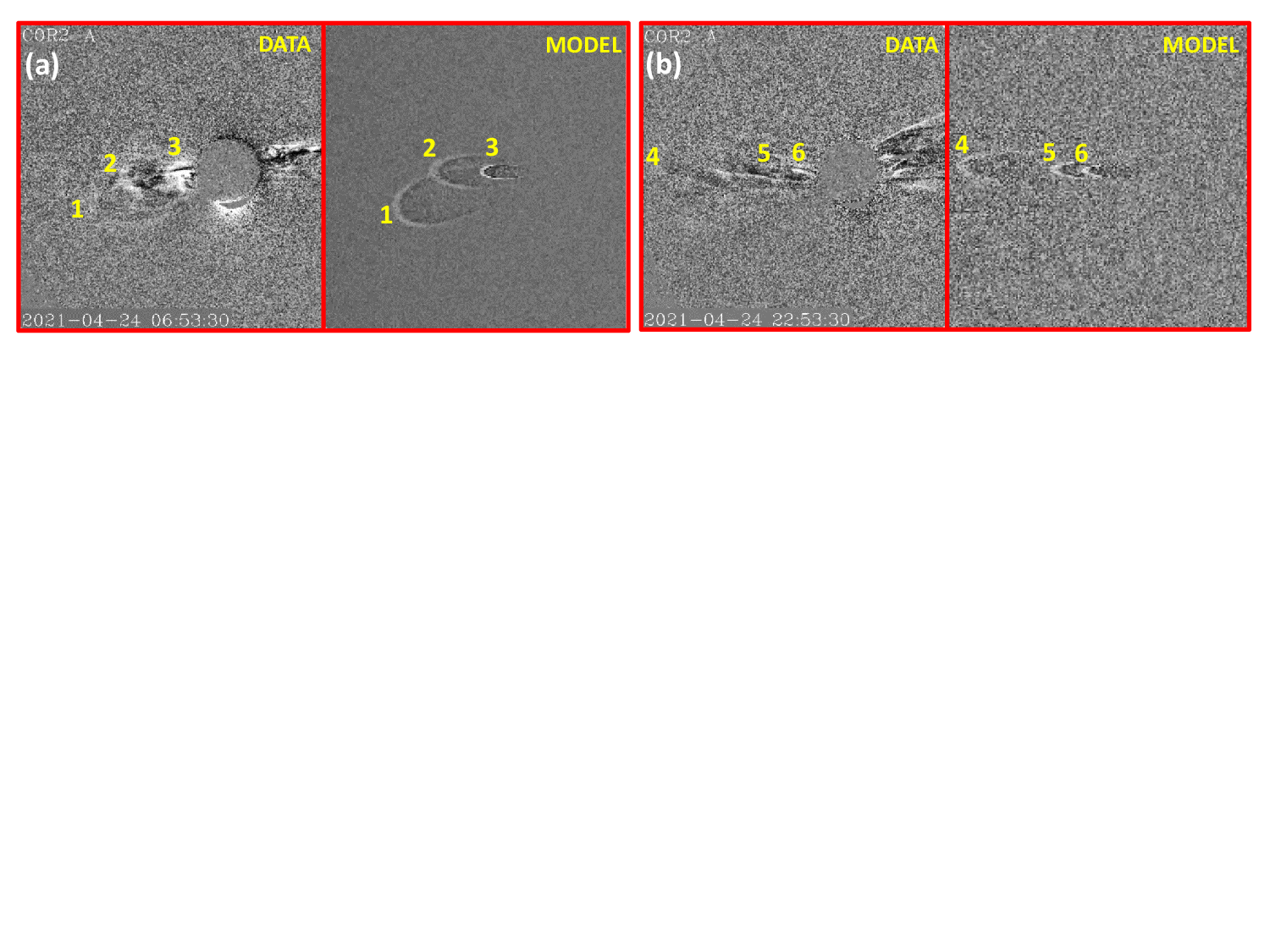}{1.5in}{0}{70}{70}{-250}{-260}
\caption{(a) On the left is an image of a series of three small CMEs
  (numbered 1-3) from UT 06:53:30 on 2021 April 24, taken by the COR2-A
  coronagraph on board STEREO-A.  To the right of the image is a synthetic
  image based on the 3-D morphological reconstruction of the CMEs described
  in Section 3.  The real and synthetic images are displayed in a
  running-difference format, and artificial noise is added to the
  synthetic image for aesthetic purposes.  (b) Another COR2-A image,
  from UT 22:53:30 on 2021 April 24, showing three later small CMEs
  (numbered 4-6), with the synthetic model image again shown next to it.
  (A movie version of this figure is available in the online
  version of this article.)}
\end{figure}
     Although the principal stimuli of our analysis are the images and in situ
observations from PSP on 2021 April 24-25, we first provide context for these
data by presenting coronagraphic observations at that time from 1 au, in
Figure 1.  These images are from the COR2-A coronagraph on the STEREO-A
spacecraft, which observes at angular distances from Sun-center of
$0.7^{\circ}-4.2^{\circ}$ ($2.5-15.6$ R$_{\odot}$) \citep{rah08}.
The images are displayed in a running-difference format,
subtracting the previous image to emphasize the dynamic CME fronts.
On 2021 April 24, COR2-A observes a series of small CMEs erupting off the
east limb of the Sun.  (The COR2-A observations of the activity are covered
more thoroughly in the movie version of Figure 1.)
This sequence of transients is not easy to separate
into distinct eruptions, but in Figure 1 we identify six separate little CMEs.
Ideally, it would be preferable if it could be determined if there are
truly separate magnetic flux rope (MFR) structures within these separate
transients, to ensure these are in fact separate eruptions, but as will
be discussed in more detail below, the available imaging data do not
allow this to be clearly determined.
This kind of sustained activity is not uncommon in coronagraphic data, and
has been frequently observed by, for example, the Large Angle and Spectrometric
COronagraph (LASCO) instrument on board the SOlar and Heliospheric
Observatory (SOHO), which has been continously observing the Sun since 1996
\citep{geb95}.  These CMEs are small enough that they could be
considered in the category of ``streamer blob'' events \citep{nrs97,ymw98}.

\begin{figure}[t]
\plotfiddle{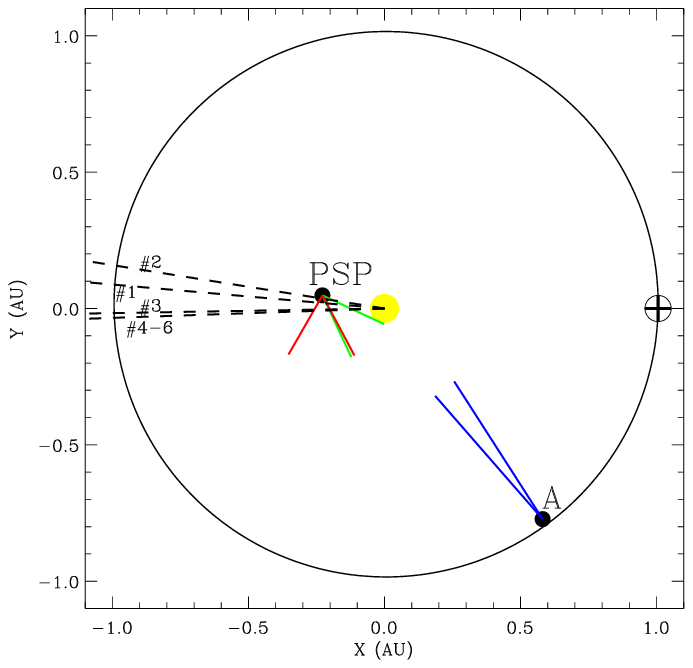}{3.6in}{0}{90}{90}{-280}{-360}
\caption{The positions of Earth, PSP, and STEREO-A in the ecliptic
  plane on 2021~April~24 (in HEE coordinates).  At PSP's position, the green and
  red lines indicate the fields of view of the WISPR-I and WISPR-O detectors.
  At STEREO-A's position, the blue line indicates the field of view of COR2-A.
  The dashed lines indicate the central trajectories of six small CMEs
  observed by all these imagers.}
\end{figure}
     Solar transients this small would normally not garner much attention,
but in this instance the eruptions are directed right at PSP,
making this our first opportunity to see what this kind of activity
looks like up close.  Figure~2 shows the locations of STEREO-A and PSP relative
to Earth on 2021 April 24.  It also explicitly shows the fields of view of
the COR2-A coronagraph, and of PSP/WISPR's two heliospheric imagers: 
WISPR-I, imaging at $13^{\circ}-53^{\circ}$ from the Sun, and WISPR-O,
imaging at $50^{\circ}-108^{\circ}$ \citep{av16}.  Dashed
lines in the figure indicate the central trajectories inferred for the six
CMEs, based on the stereoscopic analysis that will be described in
Section 3.  The CMEs all pass very close to PSP, which is almost directly
behind the Sun relative to Earth, explaining why SOHO/LASCO operating near
Earth is unable to see any of this activity.  As for STEREO-A imagery,
the CMEs are observed by COR2-A, as shown in Figure 1, but they are
too faint to be discerned closer to the Sun by the COR1-A coronagraph;
and the heliospheric imagers, HI1-A and HI2-A, are observing the
wrong side of the Sun to view them.  Thus, from STEREO-A we only have the
COR2-A data.  As noted in Section~1, there is a prominence
eruption associated with this activity apparent in EUV imagery, but we
refer the reader to \citet{tn23} for a detailed analysis of those data,
along with a discussion of potential signatures of the prominence in
the PSP in~situ data.  We do note here that we see no clear
evidence of the prominence in the white light images, which are naturally
observing farther from the Sun than the EUV images.

\begin{figure}[t]
\plotfiddle{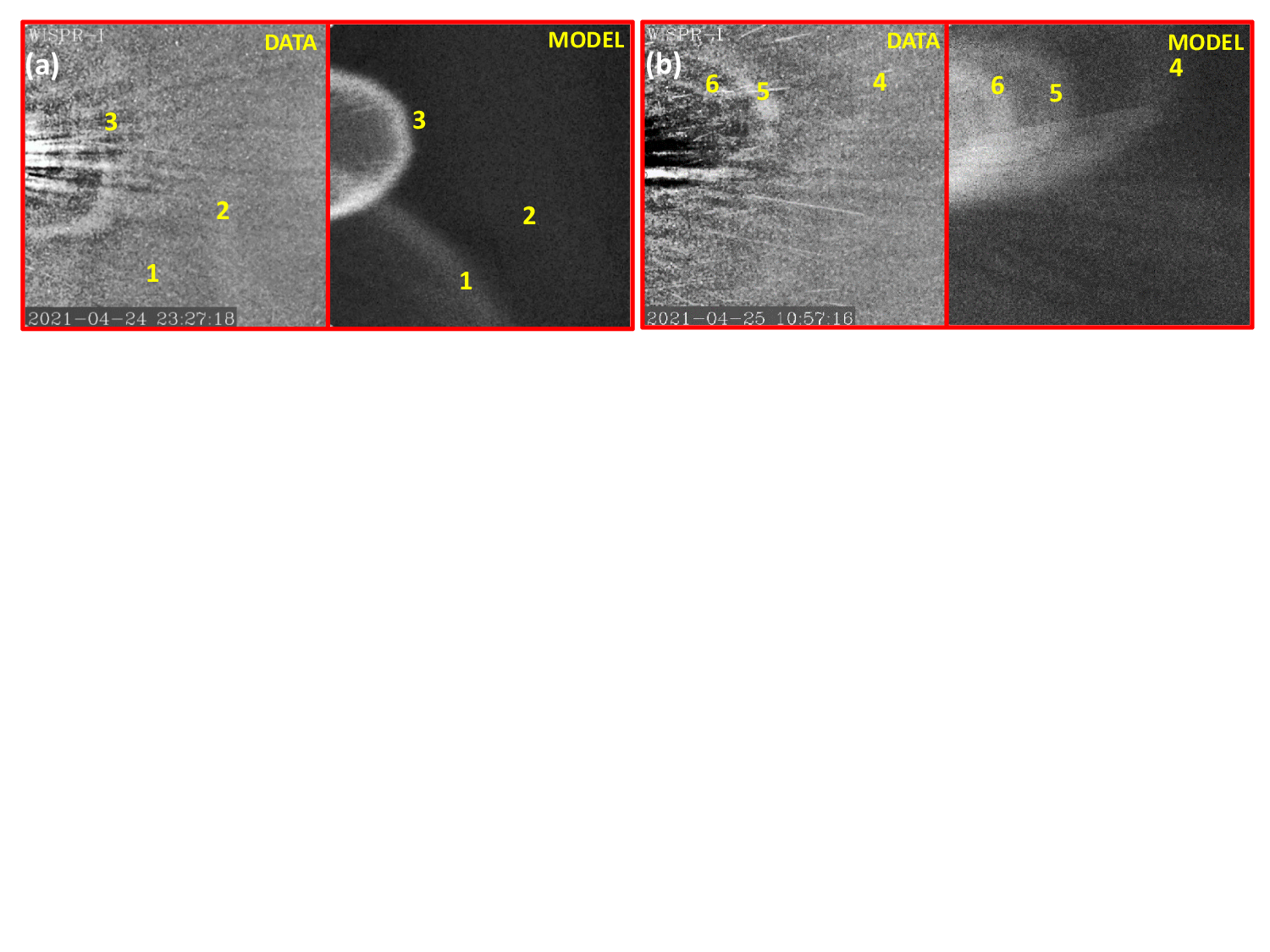}{1.5in}{0}{70}{70}{-250}{-260}
\caption{(a) On the left is a PSP/WISPR-I image of a series of
  fronts (numbered 1-3) observed on UT 23:27:18 on 2021 April 24,
  associated with the three CMEs observed by COR2-A in Figure~1(a).
  To the right of the image is a synthetic image based on the 3-D
  morphological reconstruction of the CMEs described
  in Section 3.  The real and synthetic images are displayed in a
  average-difference format, and artificial noise is added to the
  synthetic image for aesthetic purposes.  (b) Another WISPR-I image,
  from UT 10:57:15 on 2021 April 25, showing three later fronts
  (numbered 4-6), corresponding to the COR2-A CMEs seen in Figure~1(b),
  with the synthetic model image again shown next to it.
  (A movie version of this figure is available in the online
  version of this article, which also includes imagery from
  WISPR-O as well as WISPR-I.)}
\end{figure}
     In the WISPR images from 2021 April 24-25, a confusing series of fronts
is observed passing through the field of view, as illustrated by two WISPR-I
images in Figure 3.  The movie version of this figure shows this more
completely, and shows the WISPR-O imagery as well.  The images are shown
in an average-difference format, with an average WISPR-I image subtracted
from the sequence.  This activity is clearly associated with the sequence of
transients seen by COR2-A in Figure 1, but connecting the WISPR-I fronts with
the individual small COR2-A CMEs is nontrivial.
The numbers in Figure 3
indicate what fronts we ultimately associate with each CME in Figure 1.
We will be referring to these CMEs using these numbers throughout the rest of
this article.

     The first transient, CME1, is noticeably south of the ecliptic in
Figure 1, and is therefore identifiable in WISPR-I as a front only seen in
the bottom third of the images from about UT 14:27--22:27.  The other
transients, starting with CME2, are all very close to the ecliptic plane.
In the WISPR-I images, CME2 is seen as a very broad, very faint front passing
rapidly through the field of view at UT 21:57--23:57, angled diagonally from
the upper left to the lower right side of the images.  As will be shown
explicitly in the 3-D reconstruction described below in Section 3, this is
what a CME front is expected to look like when directly hitting the spacecraft,
and it is CME2 that we infer to be the first to impact PSP.  The third
transient, CME3, is the brightest in COR2-A, as shown most clearly
in the movie version of Figure 1.  We connect it with the bright,
semicircular WISPR-I front seen clearly in Figure 3, which is the first of
a series of similar looking fronts seen by WISPR-I, the most distinct of
which we connect with the CME4-CME6 sequence of transients, which are
particularly small and jet-like in COR2-A.

     We have not described the WISPR-O images much (see movie version of
Figure 3), as these data are even harder to interpret than the WISPR-I
images.  The detailed reconstruction of the eruptions described in the
next section provides some clarity.  From this it is clear that CME1 is
not seen by WISPR-O at all, being too far south; CME2 is responsible for
a bright, very broad front passing through the field of view at the end of
April 24, oriented diagnonally through the field of view (as in WISPR-I);
CME3 is a large semi-circular front filling the field of view and passing
rapidly through it from about UT 4:34--6:34 on April 25; and the final
three CME4--CME6 transients are only partially seen along the top of the
WISPR-O field of view starting at about UT 11:34, and are hard to separate.

\section{Three Dimensional Reconstruction of the Six CMEs}

     Modeling a CME's appearance in images requires both a morphological
analysis, where the 3-D shape of the CME is approximated; and a kinematic
analysis, where the Sun-center distance and speed of the CME's leading edge
as a function of time are determined.  We first discuss the kinematic
component.  For each CME, this begins with measurements of the elongation
angle of the leading edge from Sun-center, $\epsilon$, which then must be
converted to actual distances from Sun-center, $r$, using some geometric
approximation.  A kinematic model can then be inferred from these $r$
values.

     In past analyses, we have usually focused on measurements from one
spacecraft, with the choice of spacecraft based on which provides the best
vantage point and/or the most extensive set of measurements covering the
widest range of distances \citep[e.g.,][]{bew17}.  However, a single
spacecraft approach does not work in this case, because we only have data
from two sources, STEREO/COR2-A and PSP/WISPR, which are observing the
CMEs at different distances from the Sun, with no overlap.  Thus, we
combine $\epsilon$ measurements from both COR2-A and WISPR, and
infer $r$ from both.  In practice, we only consider WISPR-I from PSP,
as for most of the CMEs WISPR-O is only seeing parts of the CME fronts,
with the leading edge not clearly in the field of view.  (The only clear
exception would be CME3.)  Even for WISPR-I, meaningful measurements are
not possible for CME1 or CME2.  For CME1, the leading edge is below the
WISPR-I field of view, and with CME2 directly hitting the spacecraft,
PSP is not able to view its leading edge well.

\begin{figure}[t]
\plotfiddle{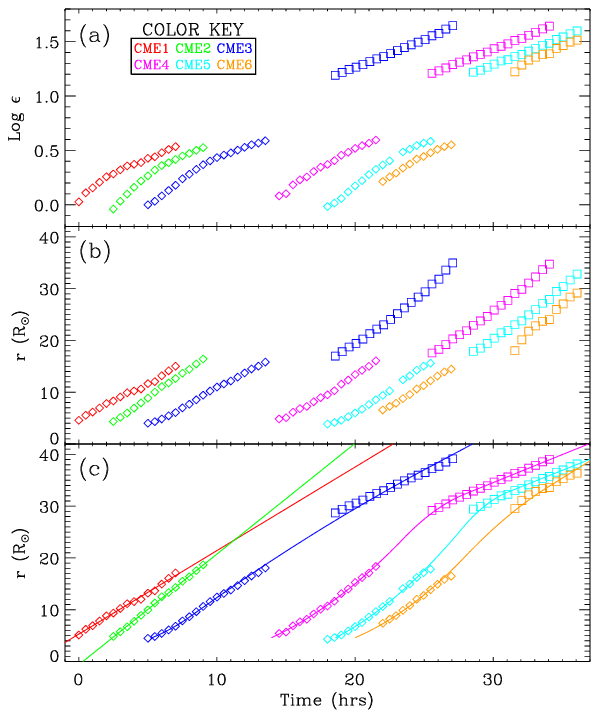}{3.7in}{0}{90}{90}{-280}{-340}
\caption{(a) Logarithmic elongation angles ($\epsilon$, in degrees) are plotted
  versus time for the leading edges of each of the six CMEs observed on
  2021 April 24--25.  The $t=0$ reference time is UT 01:23:30 on April 24.
  Diamonds are STEREO COR2-A measurements, and squares are from PSP/WISPR-I.
  (b) Sun-center distances, $r$, computed from $\epsilon$ assuming the
  Harmonic Mean approximation.  Note the severe discontinuity
  between the COR2-A and WISPR-I measurements for CME3--CME6. (c) Sun-center
  distances computed assuming the Fixed-$\phi$ approximation, with better
  COR2-A/WISPR-I agreement.  Simple kinematic models are fitted to the data
  (see text).}
\end{figure}
     The elongation angle measurements are shown in Figure 4(a) for the six
CMEs.  Converting $\epsilon$ to $r$ requires a geometric approximation.
The simplest one that is commonly used is the ``Fixed-$\phi$''
approximation, which assumes the CMEs are narrow enough for their leading
edges to be considered point-like, in which case
\begin{equation}
r=\frac{d\sin \epsilon}{\sin(\epsilon+\phi)},
\end{equation}
where $d$ is the distance from the observer to the Sun and $\phi$ is
the angle between the CME trajectory and the observer's line of sight
to the Sun \citep{swk07,nrs08}.  Alternatively,
if the CME front is approximated as a sphere centered halfway between the
Sun and the leading edge, one finds
\begin{equation}
r=\frac{2d\sin \epsilon}{1+\sin(\epsilon+\phi)},
\end{equation}
which has been called the ``Harmonic Mean'' approximation
\citep{nl09}.  More complicated relations exist
\citep[e.g.,][]{jad13}, but we will only consider the Fixed-$\phi$
and Harmonic Mean assumptions here.  It should be noted that the
trajectory directions quantified by $\phi$ and shown explicitly in
Figure 2 are determined by the morphological part of the analysis
described below.  In practice, the generation of a full CME reconstruction
involves iteration between the kinematic and morphological components of
the analysis.

     Figures 4(b) and 4(c) show the $r$ measurements that result when
the Harmonic Mean and Fixed-$\phi$ approximations are used, respectively.
It is immediately clear that the results are poor for the Harmonic
Mean option.  The distances inferred from the WISPR measurements
are too low relative to those inferred from COR2-A.
The situation is improved
when Fixed-$\phi$ is assumed.  Although a more modest discontinuity
still clearly exists, the implied kinematic profile seems more
reasonable, with the higher inferred WISPR distances.

     In a survey of 28 Earth-directed CMEs well observed by STEREO,
both Harmonic Mean and Fixed-$\phi$ were considered \citep{bew17}, and
in all but four cases it was the Harmonic Mean approximation that was
deemed better at providing a reasonable kinematic profile
that yielded correct Earth arrival times
and synthetic images that collectively best matched the multi-viewpoint
data.  However, those CMEs are bigger events than the narrow
transients being modeled here.  Because the 2021 April 24 CMEs are small and
narrow, it is ultimately not surprising that Fixed-$\phi$ does better in
this particular instance, given that the Fixed-$\phi$ approximation
assumes a very narrow transient morphology.

     Following past analyses \citep[e.g.][]{bew17},
we fit a simple multi-phase kinematic model to
the $r$ measurements in Figure 4(c), with each phase corresponding to either
a time of constant velocity or constant acceleration.  Full kinematic models
such as this allow synthetic images of CMEs to be computed for any time
whatsoever after CME initiation, including times without $r$ measurements
(e.g., WISPR-O image times, or WISPR-I image times for CME1 and CME2).
The resulting fits are represented as solid lines in the figure.  For CME1
and CME2, where we only have COR2-A measurements, we simply use a single
phase model assuming a constant velocity.  It is fortunate that this
ultimately leads to synthetic images that reproduce the actual images
reasonably well (see below), which it should be mentioned would not be the
case for CME3--CME6, which require the WISPR $r$ constraints.

\begin{table}[t]
\small
\begin{center}
Table 1:  CME Parameters
\begin{tabular}{cccccccc} \hline \hline
Parameter & Description & CME1 & CME2 & CME3 & CME4 & CME5 & CME6 \\
\hline
V$_{COR2A}$ (km/s)& COR2-A speed   & 312 & 415 & 323 & 363 & 376 & 366 \\
V$_{WISPR}$ (km/s)& WISPR-I speed  & ... & ... & 233 & 222 & 225 & 274 \\
$\lambda_s$ (deg)& Trajectory longitude & 175 & 171 & 181 & 182 & 182 & 182 \\
$\beta_s$ (deg)  & Trajectory latitude  & -12 &   0 &   0 &   1 &   1 &   1 \\
FWHM$_s$ (deg)   & Angular width        &  29 &  25 &  25 &  15 &  15 &  15 \\
$\alpha_s$       & Leading edge shape   & 2.5 & 2.0 & 2.0 & 2.0 & 2.0 & 2.0 \\
\hline
\end{tabular}
\end{center}
\end{table}
     For CME3--CME6, we assume a three-phase model, with a
period of constant acceleration followed by a period of constant
deceleration, and finally a period of constant velocity.  We have
used this kind of model for many past CMEs \citep[e.g.,][]{bew09a},
in general representing the acceleration phase of a fast CME, a gradual
deceleration as the CME is slowed by interaction with the slower ambient
wind, and finally a constant velocity as the terminal CME speed is
reached.  In this case, the meaning of the
inferred acceleration and deceleration phases is more ambiguous, as the
fit will be using the acceleration to try to account for the
aforementioned discontinuity between the COR2-A and WISPR-I
measurements.  For this reason, we do not bother to show the velocity
profiles that correspond to the fits in Figure 4(c).

However, in the first two rows of Table 1, we list mean velocities
observed for each of the CMEs in the COR2-A and WISPR-I fields of view,
computed from simple linear fits to the COR2-A and WISPR-I
height-vs-time data points.  It is
noteworthy that for CME3--CME6 we infer significantly lower velocities in
WISPR-I than for COR2-A.  Although systematic uncertainties in this
analysis are high, we believe these decelerations are probably real,
at least qualitatively, as they can be explained as a natural consequence
of these later CMEs plowing into the slower trailing parts of CME1 and CME2.

     Turning our attention to the morphological part of the analysis, we
rely on techniques used many times in the analysis of stereoscopic CME
imaging, dating back to \citet{bew09a}, but more thoroughly described
by \citet{bew17}.  This is a forward modeling approach, involving the
use of a parametrized functional form for the assumed CME shape, with
the parameters adjusted to yield synthetic images that best match the
observations.  Synthetic image calculation includes full computation of the
Thomson scattering from the model structure, although we only place mass on
the surface of the structure to outline its shape.  We are not
here attempting to reproduce image brightnesses or measure CME masses, so
the densities assumed at the surface are arbitrary.  A popular paradigm
for CME structure is that of a magnetic flux rope (MFR), i.e.\ a tube shape
permeated with a helical magnetic field with both legs streching back
toward the Sun, so we have usually assumed an MFR shape in 3-D CME
reconstructions \citep{bew09a}.  However, for the small,
faint CMEs studied here,
we really only perceive the transients as narrow fronts, with nothing to
suggest the presence or orientation of an MFR.  Thus, we instead
assume here a simple lobular front shape, a prescription we
have generally used in the past to model CME shocks, when visible ahead
of the ejecta \citep[e.g.,][]{bew09a,bew17}, but
which has also been used before to model the ejecta itself in other
cases where an MFR shape is not clearly apparent \citep{bew09b}.

\begin{figure}[t]
\plotfiddle{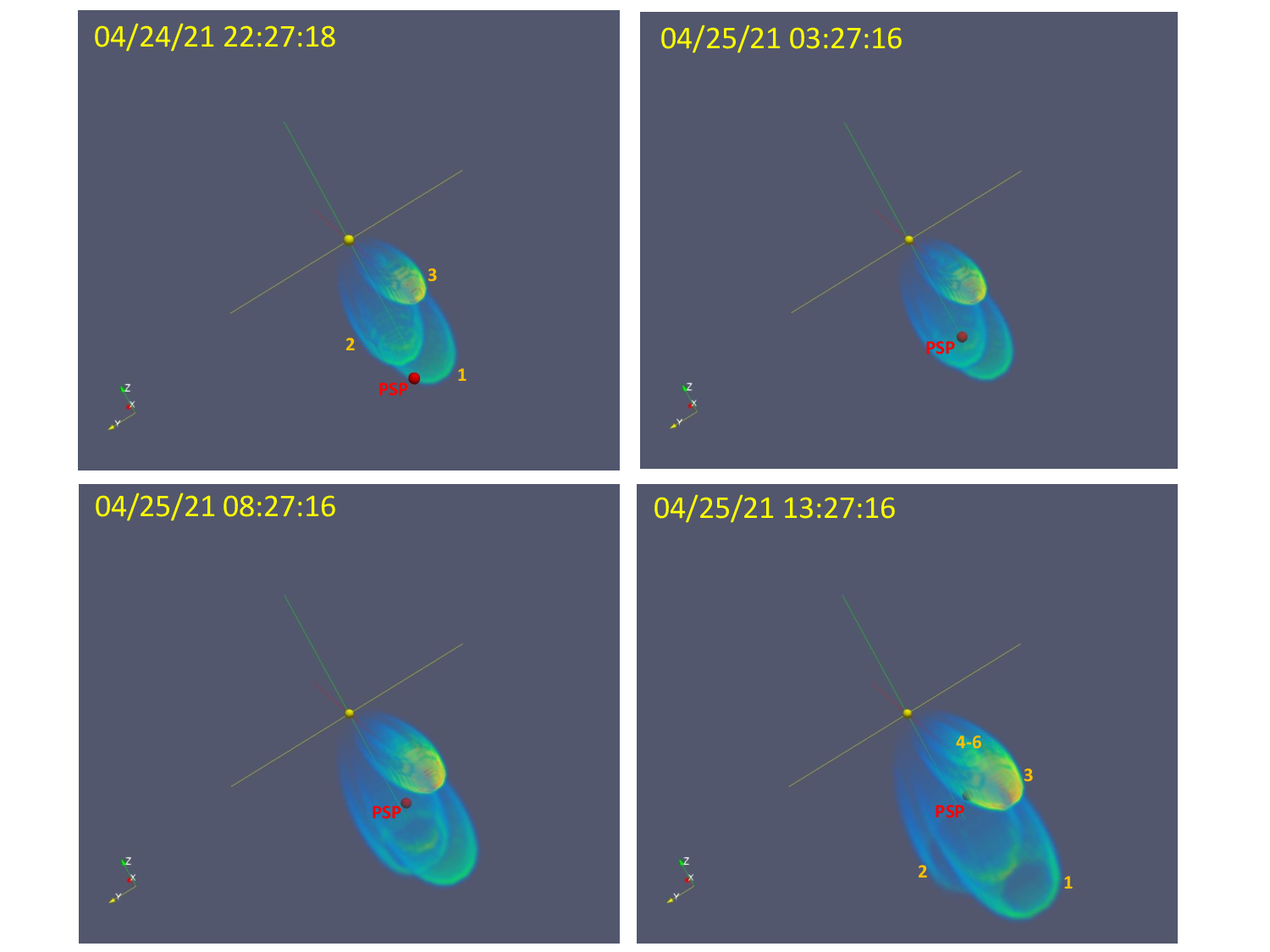}{3.6in}{0}{56}{56}{-200}{-10}
\caption{Three dimensional visualizations of the six reconstructed CME
  fronts, for four different times.  The fronts are displayed in an
  HEE coordinate system, with the size of the Sun shown to scale in
  the figures.  The time-dependent location of PSP is shown as a small
  red sphere.  The first two frames show the positions of the first
  3 CMEs (CME1--CME3) before and after CME2 hits PSP, respectively.
  The final frame shows PSP entering the flank of CME3.  The last
  3 very narrow CMEs (CME4-CME6) are buried inside CME3, as labeled
  in the final frame.}
\end{figure}
     Figure 5 shows the resulting model CME fronts, and their positions
relative to each other and to PSP.  The parameters used to define the
fronts are listed in the last four rows of Table 1.  The trajectory
longitude and latitude of the CMEs are $\lambda_s$ and $\beta_s$,
respectively, based on a heliocentric Earth ecliptic (HEE) coordinate
system, with the positive x-axis pointed toward Earth, and the positive
z-axis pointed toward ecliptic north.  The CME longitudes are shown
explicitly in Figure~2, which are within
$11^{\circ}$ of each other.  Only CME1 has a
latitude significantly out of the ecliptic ($\beta_s=-12^{\circ}$).
The $\phi$ trajectory angles relative
to STEREO-A and PSP used in equations (1) and (2) above can be
computed from $\lambda_s$, $\beta_s$, and the known
spacecraft positions.  The $\alpha_s$ parameter in Table 1 defines
the shape of the leading edge, with higher values of $\alpha_s$
being flatter and lower values being more rounded
\citep[see, e.g.,][]{bew17}.

     The widths of the CME fronts are quantified in Table 1
as a full-width-at-half-max, $FWHM_s$.  The CMEs are
all very narrow, with $FWHM_s$ ranging from $15^{\circ}$ for
CME4--CME6 to $29^{\circ}$ for CME1.  These values can be crudely
compared with the widths inferred for CMEs modeled with MFRs,
with $FWHM_s$ in those cases defining the width of the CME in the
plane of the MFR.  In the survey of 28 Earth-directed CMEs of
\citet{bew17}, the narrowest CME had $FWHM_s=41.5^{\circ}$.
Even the tiny streamer blob CME observed by PSP/WISPR in
2018 November 5 had $FWHM_s=43.6^{\circ}$ \citep{bew20}.
All this emphasizes that we are not perceiving much lateral
extent for the small 2021 April 24 CMEs
compared to past CMEs that we have studied stereoscopically.

     The CME trajectories and morphological parameters were determined
by a trial and error process of experimenting with different parameters
to see which parameters lead to synthetic images that best match the
real data.  The self-similar expansion of the model CME fronts is
defined by the kinematic models in Figure 4(c).  The synthetic images
resulting from our best-fit models are compared with the real images
in Figures 1 and 3, with the movie versions of these figures available
in the online article providing a more thorough comparison.  Artificial
noise is added to the synthetic images for aesthetic purposes.
The added noise level is the same in absolute terms in all the images.
The apparent lower noise in Figure 1(a) compared to 1(b) is illusory
due to the upper range of the color scale being extended upwards to
match the range of CME3, which is the brightest of the six CMEs.

     It is difficult to estimate uncertainties in the
parameters listed in Table~1.  There have been many attempts to
estimate errors in CME reconstructions based on multi-spacecraft
imaging.  A recent example is \citet{cv23}, which also summarizes much
of the past work on the subject.  A lot of this work focuses on
kinematic uncertainties, due to the interest in predicting 1~au
arrival times at Earth for CMEs
\citep[e.g.,][]{ekjk12,mlm15,bew17,amw18,ep21}, but
morphological uncertainties are studied as well
\citep[e.g.,][]{at09,mm10,sj16,lab18}.  Uncertainties in an analysis
of this nature will depend greatly on the characteristics of the
particular event in question, and on the particular viewing
geometry for that event.  The applicability of these studies to
the very unique viewing geometry of the 2021 April 24 events directed
right at PSP is unclear.

     One characteristic of our reconstruction that is
worthy of note is that the synthetic WISPR images are remarkably
sensitive to the assumed trajectory direction, with shifts of only
a degree or two making noticeable changes to the images.  This is
simply due to the extreme close proximity of the CMEs to the
spacecraft (see Figure 2).  Perhaps the best examples of this are the
latitude parameters for CME3--CME6.  For CME3, we infer
$\beta_s=0^{\circ}$, while for CME4--CME6 we infer $\beta_s=1^{\circ}$.
Such a small difference would usually not be significant in a
CME reconstruction of this nature, but in this case it is significant.
This is best indicated by the WISPR-O data, where
the CME3 front is fully visible in the WISPR-O field of view, but the
CME4--CME6 fronts are at least halfway above the field of view.
Only a single degree of latitude shift in the reconstruction is
enough to reproduce this.

     With the image-based CME reconstruction now complete, we can finally
address the central question of which CMEs are responsible for the
in situ signatures observed by PSP.  Our conclusions are illustrated by
the four time frames shown in Figure 5.  The first CME, CME1,
is directed south of the ecliptic ($\beta_s=-12^{\circ}$) and therefore
does not hit PSP.  It is CME2 that makes a direct hit to PSP at the end of
April 24.  The final panel of Figure 5 then shows PSP encountering the
flank of CME3, with PSP's orbital motion being partly responsible for
carrying the spacecraft into CME3, in addition to the expansion of
the CME itself.  The last three CMEs, CME4-6, are highly blended by
the time they reach the vicinity of PSP.  It is likely that PSP encounters
trailing parts of these little transients near the beginning of April 26.

\section{Comparison with PSP In-Situ Data}

\begin{figure}[t]
\plotfiddle{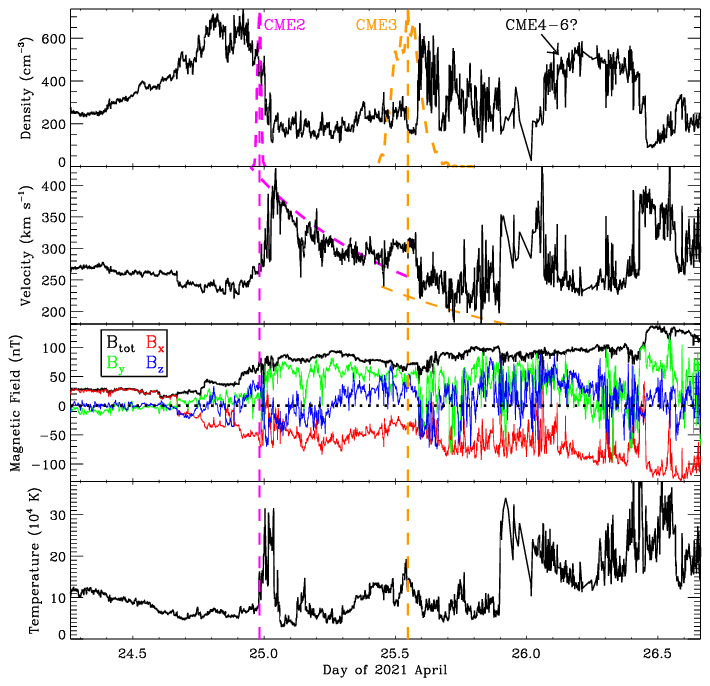}{3.7in}{0}{95}{95}{-280}{-380}
\caption{Plasma and field measurements from PSP/SWEAP and PSP/FIELDS
  during the passage of the 2021 April 24 transients over the
  spacecraft.  The top two panels are proton number density and
  velocity.  The third panel is the magnetic field at PSP, both
  the total field and the $B_x$, $B_y$, and $B_z$ coordinates in
  an RTN coordinate system.  The final panel is the proton
  temperature.  Vertical dashed lines indicate the arrival times
  of CME2 and CME3 predicted by the 3-D reconstruction in Figure 5,
  and the density and velocity profiles predicted by this reconstruction
  are explicitly shown as well.  The velocity profiles match the
  data well.  The 3-D reconstruction implies that the broad density
  increase early in April 26 may be due to the last 3 modeled transients,
  CME4-6.}
\end{figure}
     We will now compare the predictions of our image-based event
reconstruction with measurements from PSP's plasma and field
instruments, in order to see if the in~situ data provide support for
the reconstruction.  The in~situ data are shown in Figure 6.  The proton
density, velocity, and temperature measurements are from the
Solar Wind Electrons Alphas \& Protons (SWEAP) instrument on PSP
\citep{jck16}, while the magnetic field measurements
are from the FIELDS instrument \citep{sdb16}.

     Vertical dashed lines in Figure 6 indicate when the 3-D
reconstruction has PSP being hit by CME2 and CME3, the first right at
the end of April 24, and the second midway through April 25.
The PSP in~situ data broadly support these predictions.  The predicted
CME2 encounter time corresponds to a substantial increase in
velocity, magnetic field, and temperature.  Proton densities actually
show a broad peak ahead of this arrival time, suggesting a pile-up
region ahead of the CME.  Possible signatures of the predicted CME3
encounter are more subtle, but there is a significant density increase
at that time, along with a velocity decrease, and a modest
temperature peak.

     We do not try to show predicted encounter times for CME4-6,
because these last three little transients are rather blended and are
predicted to have an extended grazing incidence encounter with PSP,
making predictions more uncertain.  However, we do note in Figure 6 the
existence of a broad density peak during the first half of April 26,
which would correspond to the time
when we would expect PSP to be affected by the passage of CME4-6.

     The density and velocity profiles predicted at PSP for CME2 and
CME3 by our 3-D reconstruction are explicitly shown in the top two
panels of Figure 6.  The predicted velocity profiles indicate
decreasing velocity behind the CME arrival, characteristic
of self-similar expansion.  These predictions agree well
with the PSP/SWEAP observations, particularly for CME2.  This
is encouraging, considering the difficulties in deriving accurate
kinematic models for these events (see Section 3).

     A precise replication of the observed density profile is not
expected, considering the very rudimentary density distribution
assumed in the reconstruction, with mass placed only on the surfaces
of the CME fronts.  Gaussian density profiles across the surface are
assumed, so the predicted PSP density profiles are roughly Gaussian as
well.  For CME2 the predicted profile is very narrow, indicative of
the direct hit of CME2 on PSP, carrying PSP quickly through the CME
surface.  In contrast, the predicted density profile for CME3 is
broad, as PSP enters CME3 with more of a grazing incidence (see last
panel of Figure 5).  

     As for the magnetic field behavior, before the transients arrive
PSP sees a positive-polarity Parker spiral behavior, with
positive $B_x$ and much weaker negative $B_y$.  This changes in the
pile-up region ahead of CME2, with $B_x$ and $B_y$ reversing
sign, and the overall field strength increasing dramatically as
CME2 arrives at the end of April 24.  However, after CME2's arrival,
the field behavior remains relatively consistent, despite the
passages of the later transients.  Throughout this period, $B_x$
remains consistently negative and $B_y$ generally positive.

     One obvious thing to do with the in situ data is to look for
evidence of MFRs, as it is generally assumed that
MFRs lie at the hearts of all CMEs \citep{av13}.  We have
already noted that for these small, relatively faint CMEs, we perceive
the transients in the white light images as simple lobular fronts,
with no clear evidence of an MFR shape, but this does not preclude the
existence of undetectable MFRs behind the visible leading fronts.
However, the in~situ data do not show obvious ``magnetic cloud''
signatures, which are considered the clearest indication of an MFR;
with smoothly rotating, strong fields, low density, and low temperature
\citep{lb81,km86,lfb88,rpl90,rpl15,vb98}.  The absence of
such a signature is particularly notable for CME2, for which our 3-D
reconstruction implies a direct hit on PSP, which we would expect to
yield an MFR signature in the in situ data, if an MFR is
in fact present.

\begin{figure}[t]
\plotfiddle{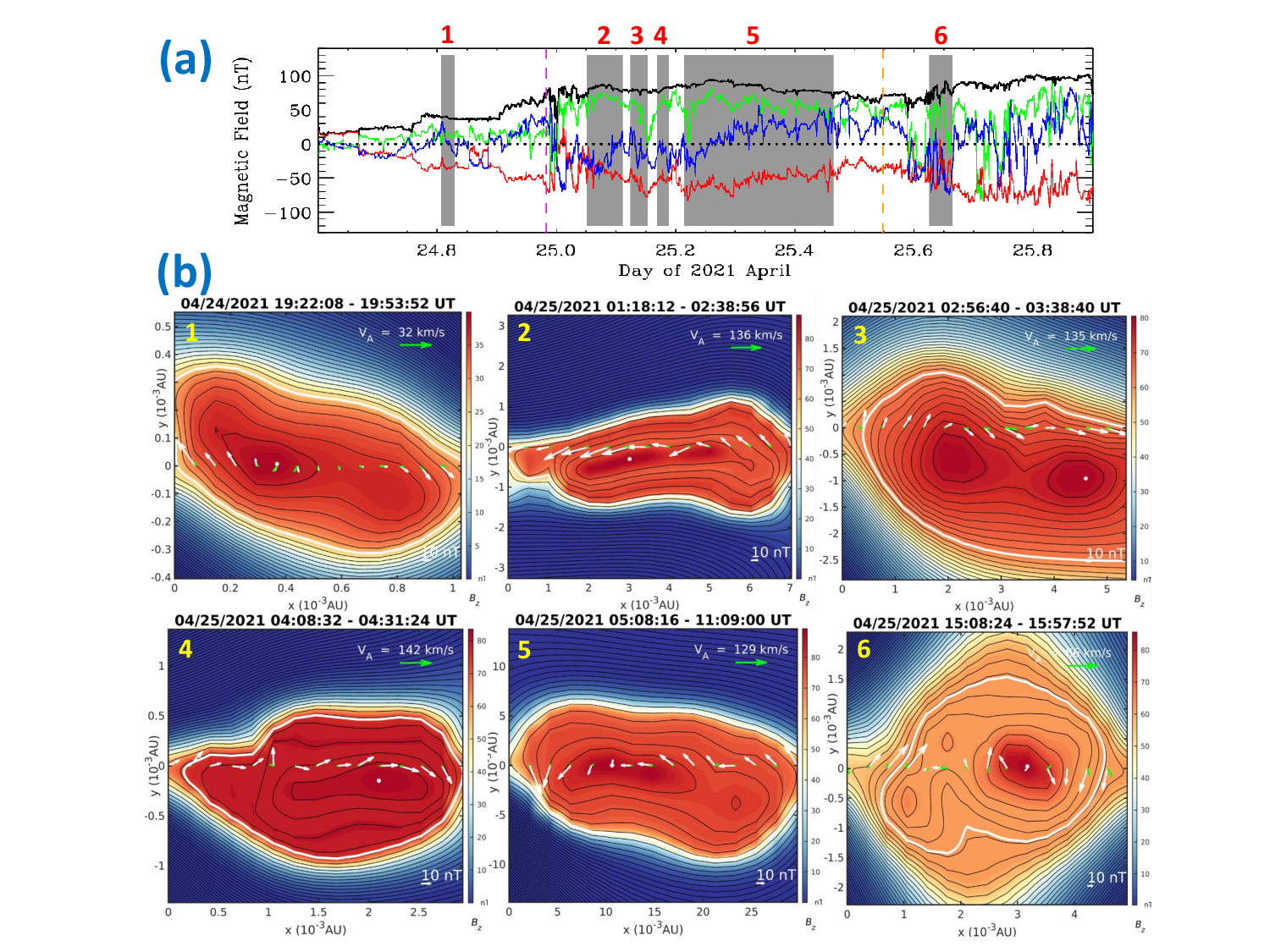}{4.0in}{0}{63}{63}{-240}{-20}
\caption{(a) The magnetic field panel from Figure~6 is reproduced,
  but with shaded regions indicating six intervals identified
  as possible MFRs by an automated routine (see Table 1).  Vertical
  dashed lines
  indicate the arrival times of CME2 and CME3, as in Figure~6.
  (b) Grad-Shafranov reconstructed cross sections of the six MFRs.
  The color scale is the out-of-plane axial field in nT.  The contours
  are the in-plane transverse magnetic field lines.  The white
  and green arrows along $y=0$ are the measured transverse field and
  remaining flow vectors, respectively, along the spacecraft path.}
\end{figure}     Nevertheless, there is field variability during the CME
encounter, which could be indicative of distinct magnetic features
that could be interpreted as MFRs.  In order to explore
this, we use an automated routine for finding MFRs within
the solar wind, developed by \citet{qh18}.  Most MFRs found by
this technique are quite small, with a median encounter duration of
20 minutes in {\em Wind} data.  The basis of the
routine is the Grad-Shafranov (GS) technique for inferring MFR
orientations and deriving two-dimensional configurations
for clear magnetic cloud events within in~situ solar
wind data \citep{qh02,qh17}.  Unlike more classic techniques
for modeling such structures \citep[e.g.,][]{rpl90,rpl15,tnc16},
the GS methodology does not
assume that the MFR has a symmetric circular or elliptical MFR
cross section.  The broader, more general theoretical requirements
of the GS approach allows the automated \citet{qh18} MFR
search routine to identify a greater number of potential MFR
candidates within the solar wind than more restrictive techniques.

\begin{table}[t]
\small
\begin{center}
Table 2:  Flux Rope Properties
\begin{tabular}{clcccccc} \hline \hline
MFR\# & Start Time & Duration & \multicolumn{2}{c}{Orientation}&
  $F_{\phi}$ & $F_{\psi}$ & $F_{\psi}/F_{\phi}$ \\
      &            &   (s)    &  $\phi$ (deg) & $\theta$ (deg) &
     ($10^{10}$ Wb)       &  ($10^{10}$ Wb)        &                  \\
\hline
1 & 2021-04-24 19:22:08 &  1904 & 160 &  80 & 0.019 & 1.87 & 96.4  \\
2 & 2021-04-25 01:18:12 &  4844 & 140 & 100 & 2.21  & 36.7 & 16.6 \\
3 & 2021-04-25 02:56:40 &  2520 & 120 & 100 & 2.09  & 39.9 & 19.1 \\
4 & 2021-04-25 04:08:32 &  1372 & 120 & 100 & 0.49  & 11.0 & 22.2 \\
5 & 2021-04-25 05:08:16 & 21644 & 140 &  80 & 61.0  & 194  & 3.18 \\
6 & 2021-04-25 15:08:24 &  2968 & 140 & 130 & 1.19  & 37.9 & 31.8 \\
\hline
\end{tabular}
\end{center}
\end{table}
     \citet{qh18} identified 74,241 small MFRs in {\em Wind}
data from 1996--2016.  \citet{yc22} applied these techniques
to PSP data from the first six PSP orbits, finding nearly 6000
MFRs.  We now similarly apply the algorithm to our time interval
of interest in 2021 April 24-26.  Focusing only on longer candidates
($\geq 20$ min), we identify six possible MFR structures for
which we can compute a full GS reconstruction.  The time intervals
of these MFRs are listed in Table~2 and displayed in Figure~7(a).
The first MFR (MFR1) is in the pile-up region ahead of CME2's arrival at
PSP based on the 3-D reconstruction.  The next four MFRs are
sequential within the time period that we associate with the CME2
encounter.  These four MFRs collectively account for most of that
time period.  The last of these (MFR5) is by far the largest, with
an encounter duration of 6.0 hr.  The final MFR (MFR6) is in the CME3
encounter period.

     The full GS reconstructions of the six MFRs are shown in Figure~7(b),
with a color scale that maps the axial field component of the MFR in the
cross-sectional plane.  In the GS modeling, the MFR is assumed to extend
infinitely out of the plane in both directions, and the figures indicate
the inferred spacecraft path through these structures along $y=0$, projected
into the cross-sectional plane.  The orientation of the MFR is indicated
by the $\phi$ and $\theta$ directions in Table~2, in the spacecraft-based
RTN coordinate system.  The $\phi$ angle indicates
the azimuthal direction, with $\phi=0^{\circ}$ and $\phi=180^{\circ}$
pointed radially away and toward the Sun, respectively; while
$\theta$ is the polar angle, with $\theta=0^{\circ}$ and
$\theta=180^{\circ}$ pointed toward ecliptic north and south, respectively.

\begin{figure}[t]
\plotfiddle{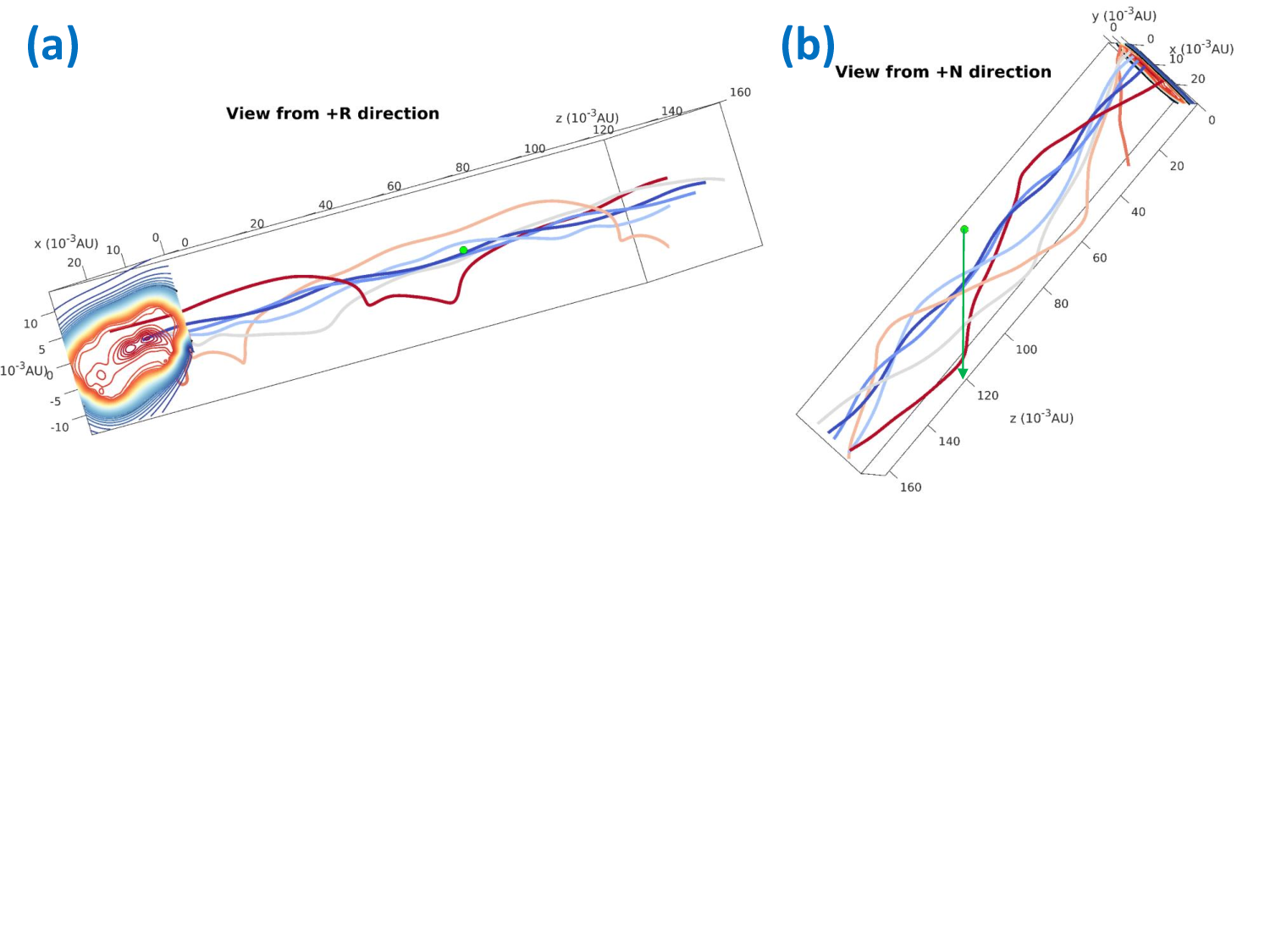}{2.7in}{0}{70}{70}{-235}{-180}
\caption{(a) The 3-D magnetic field configuration of MFR5 (see Figure 7),
  viewed from a positive-R direction, looking toward the Sun, with the
  green dot indicating PSP's path through the structure.  The field lines are
  traced from the cross-sectional map on the left.  (b) Here MFR5 is viewed
  from a positive-N direction, looking down on the MFR with the Sun toward
  the bottom, and the green arrow indicating PSP's path through the MFR.}
\end{figure}
     With $\theta=80^{\circ}-100^{\circ}$, the five CME2-associated MFRs
are all oriented close to the ecliptic plane.
With $\phi=120^{\circ}-160^{\circ}$, they are all encountered with the right
side of the MFR slanted toward the Sun relative to the spacecraft-Sun line.
Figure~8 explicitly shows the orientation of MFR5, with panel (b) showing
how the MFR is tilted relative to the spacecraft's path in the ecliptic plane.
With actual MFR structures assumed to have legs that stretch back
toward the Sun, the inferred [$\phi$,$\theta$] angles could be consistent
with PSP encountering the right sides of roughly E-W oriented MFRs.
However, it should be noted that the GS analysis assumes that the MFR
extends infinitely in directions out of the cross-sectional plane.  This
introduces significant uncertainty in the interpretation of the MFR
orientation angles on large scales.  In the example of the large CME2
MFR (e.g., MFR5), significant curvature would be expected for it to be
contained within the confines of the rather narrow CME outline defined
by the CME lobular front inferred from the 3-D reconstruction.

     Besides the MFR orientation information, Table~2 also provides
estimates of the toroidal and poloidal magnetic fluxes of the
GS-constructed MFRs, $F_{\phi}$ and $F_{\psi}$, respectively.
The toroidal or axial flux is easiest to understand,
as it is simply the magnetic flux through the cross-sectional
plane illustrated in Figure~7 for each MFR.  The poloidal flux is the
integration of the azimuthal field around the central axis times the
area of the surface through which this field is passing.  This flux
is much more uncertain than $F_{\phi}$, since one must first
assume that the azimuthal field times the MFR radius is unchanged
along the MFR length (i.e., under a 2-D geometry),
and also a total length of the MFR must be
known or assumed.  In principle, this could come from an image-based
reconstruction of the MFR shape, but in this case the MFR shapes are
not discerned in the images and we can only infer an outline of the
CME leading edge.  Still, we can use the length of the curve outlining
the lobular fronts as our estimate for MFR length.  At a distance from
the Sun of 45~R$_{\odot}$, this length is 94.6~R$_{\odot}$ based on the
CME2 front shown in Figure~5, so the $F_{\psi}$ values in Table~2 are
computed assuming this length.  Finally, the last column of Table~2
lists the ratio of $F_{\psi}$ to $F_{\phi}$.  These ratios are higher
for the smaller MFRs.

\section{Summary and Discussion}

     The 2021 April 24-25 event is the first CME that directly
hits PSP and is simultaneously imaged by it.  We have performed
a full 3-D reconstruction of the event based on the available
imagery, which unfortunately is limited to COR2-A and WISPR.
The reconstruction is also complicated because the activity
actually consists of multiple little CMEs erupting sequentially
from the solar corona, but we have successfully connected six
distinct, small CMEs seen by COR2-A near the Sun with fronts
seen by WISPR-I and WISPR-O as the transients pass either over or
very near PSP.  The analysis clearly identifies CME2 as the
transient that first impacts PSP directly, with PSP later
predicted to encounter CME3 more obliquely.

     We have found support for the 3-D reconstruction from
PSP in~situ data, with density, temperature, and field
signatures of the expected CME2 and CME3
impacts present in the in~situ data.  The declining velocity
profiles observed in~situ are also in excellent agreement with
those predicted by the reconstruction, which simply
assumes self-similar expansion.  A broad density enhancement
early in April~26 may be due to the collective influence of CME4-6.

     Finally, we have used the GS-based automated MFR finding
routine of \citet{qh18} to identify MFR candidates within
the PSP encounter time, ultimately focusing on six of the longer
candidates, all but the last associated with the CME2 encounter period.
The fifth MFR is by far the largest, with an encounter duration of
about 6 hours.  The five CME2 MFRs have orientations roughly consistent
with PSP encountering the right sides of roughly E-W oriented MFRs,
which are sloping back towards the Sun.

     The multiple eruptions seen by COR2-A are suggestive of
intermittent reconnection occuring near the heliospheric current
sheet, an interpretation consistent with the PSP in~situ data showing
a clear change in field polarity at the beginning of the CME
encounter.  Small-scale magnetic islands are commonly observed
near current sheets \citep{ok15}, which could be
interpreted as small MFR structures, and which are potential sources
of particle acceleration \citep{gpz17,kve22}.
After initial formation, small MFRs are expected to merge into
larger ones \citep{jfd21}.  It is
possible that the sequence of CME2 MFRs in Figure~7 is indicative
of such a process, with a large, dominant MFR (MFR5) preceded by
much smaller MFRs that never merged with the large MFR before
flowing away from the reconnection site.

     It is becoming more popular to envision microscale interchange
reconnection as being responsible for both the ubiquitous
switchback structures observed by PSP \citep{jfd21}, and
potentially for the solar wind as a whole \citep{ymw20,ner23}.
Activity such as the 2021 April 24-25 events
may provide a way to study this important process on a more
macroscopic scale, large enough for the intermittent outflows from
such reconnection to be apparent in imaging.  The 2021 April 24-25
events are potentially particularly useful for such purposes,
with PSP providing both imaging and in~situ observations
close to the Sun.  Thus, these data may be worthy of more detailed
analysis in the future.

\acknowledgments

     Financial support was provided by the Office of Naval Research.
QH and YC acknowledge support from NASA through grants
80NSSC21K1763 and 80NSSC21K0003, and from NSF through award AGS-2229065.     
Parker Solar Probe was designed, built, and is now operated by the
Johns Hopkins Applied Physics Laboratory as part of NASA's Living with a
Star (LWS) program (contract NNN06AA01C).  We particularly acknowledge
the WISPR instrument team, funded by NASA through grant NNG11EK11I, the
SWEAP team led by J.\ Kasper, and the FIELDS experiment led by S.\ Bale.
The STEREO/SECCHI data are produced by a
consortium of NRL (US), LMSAL (US), NASA/GSFC (US), RAL (UK), UBHAM
(UK), MPS (Germany), CSL (Belgium), IOTA (France), and IAS (France).
In addition to funding by NASA, NRL also received support from the
USAF Space Test Program and ONR.

\end{document}